\documentclass[11pt]{article}

\usepackage{fullpage}
\usepackage{color}

\usepackage{amsmath}
\usepackage{graphics}
\usepackage{subfigure}
\usepackage{multirow}
\usepackage{wrapfig}
\usepackage{graphicx}
\usepackage{verbatim}
\usepackage{amsfonts}
\usepackage{algorithm}
\usepackage{algorithmic}
\usepackage{array}
\usepackage{listings}
\usepackage{url}
\usepackage{caption}

\definecolor{mygreen}{rgb}{0,0.6,0}
\definecolor{mygray}{rgb}{0.5,0.5,0.5}
\definecolor{mymauve}{rgb}{0.58,0,0.82}

\title{Teaching Concurrent Software Design: \\ A Case Study Using Android}
\author{Konstantin L\"aufer and George K. Thiruvathukal\\
Loyola University Chicago\\
\{laufer,gkt\}@cs.luc.edu\\
}

\begin{document}

\maketitle

\begin{abstract}
In this article, we explore various parallel and distributed computing topics from a user-centric software engineering perspective. 
Specifically, in the context of mobile application development, we study the basic building blocks of interactive applications in the form of events, timers, and asynchronous activities, along with related software modeling, architecture, and design topics.
\end{abstract}

\begin{description}
\item[Relevant software engineering topics:]
software requirements: functional requirements (C), nonfunctional requirements (C)
software design: user interface patterns (A), concurrency patterns (A), testing patterns (A), architectural patterns (C), dependency injection (C), design complexity (C);
software testing: unit testing (A), managing dependencies in testing (A);
cross-cutting topics: web services (C), pervasive and mobile computing (A)
\item[Relevant parallel and distributed computing topics:]
Algorithmic problems: asynchrony (C);
architecture classes: simultaneous multithreading (K), SMP (K);
parallel programming para\-digms and notations: task/thread spawning (A);
semantics and correctness issues: tasks and threads (C), synchronization (A);
concurrency defects: deadlocks (C), thread safety/race conditions (A);
cross-cutting topics: why and what is parallel/distributed computing (C), concurrency (A), nondeterminism (C)
\item[Learning outcomes:]
The student will be able to model and design mobile applications involving events, timers, and asynchronous activities. 
The student will be able to implement these types of applications on the Android platform.
The student will develop an understanding of nonfunctional requirements.
\item[Context for use:]
A semester-long intermediate to advanced undergraduate course on object-oriented development.
Assumes prerequisite CS2 and background in an object-oriented language such as Java, \verb|C++|, or \verb|C#|.
\end{description}

\setlength\marginparwidth{5em}
\newcommand\todo[1]{\marginpar{\textbf{TODO:} \emph{#1}}}



\section{Background and Motivation}

In this article, we will explore various parallel and distributed computing topics from a user-centric software engineering perspective. 
Specifically, in the context of mobile application development, we will study the basic building blocks of interactive applications in the form of events, timers, and asynchronous activities, along with related software modeling, architecture, and design topics. 

Based on the authors' ongoing research and teaching in this area, this material is suitable for a five-week module on concurrency topics within a semester-long intermediate to advanced undergraduate course on object-oriented development. 
It is possible to extend coverage by going into more depth on the online examples~\cite{CS313Ex} and studying techniques for offloading tasks to the cloud~\cite{CS313PrimeChecker}.
The article is intended to be useful to instructors and students alike.

Given the central importance of the human-computer interface for
enabling humans to use computers effectively, this area has received
considerable attention since around 1960~\cite{Myers1998}. Graphical
user interfaces (GUIs) emerged in the early 1970s and have become a
prominent technical domain addressed by numerous widget toolkits
(application frameworks for GUI development). Common to most of these
is the need to balance ease of programming, correctness, performance,
and consistency of look-and-feel. Concurrency always plays at least an
implicit role and usually becomes an explicit programmer concern when
the application involves processor-bound, potentially long-running
activities controlled by the GUI. Here, long-running means anything
longer than the user wants to wait for before being able to continue
interacting with the application. This article is about the concepts
and techniques required to achieve this balance between correctness
and performance in the context of GUI development.

During the last few years, mobile devices such as smartphones and
tablets have displaced the desktop PC as the predominant front-end
interface to information and computing resources. 
In terms of global internet consumption (minutes per day), 
mobile devices overtook desktop
computers in mid-2014~\cite{Digiday2016}, and 
``more websites are now loaded on smartphones and tablets than on
desktop computers''~\cite{Guardian2016} as of October 2016.
Google also announced~\cite{Googleblog2015} that it will be displaying mobile-friendly web sites higher in the search results, which speaks to the new world order.
These mobile devices participate in a massive global distributed
system where mobile applications offload substantial resource needs
(computation and storage) to the cloud.

In response to this important trend, 
this article focuses on concurrency in the context
of mobile application development, especially Android, which shares
many aspects with previous-generation (and desktop-centric) GUI
application frameworks such as Java AWT and Swing yet. (And it almost
goes without saying that students are more excited about learning
programming principles via technologies like Android and iOS, which
they are using more often than their desktop computers.)

While the focus of this article is largely on concurrency within the
mobile device itself, the online source code for one of our examples~\cite{CS313PrimeChecker} goes
beyond the on-device experience by providing versions that connect to
RESTful web services (optionally hosted in the
cloud)~\cite{Christensen:2009}. We've deliberately focused this article
around the on-device experience, consistent with ``mobile first''
thinking, which more generally is the way the ``internet of things''
also works~\cite{Ashton2009}. This thinking results in proper
separation of concerns when it comes to the user experience, local
computation, and remote interactions (mediated using web services).


It is worth taking a few moments to ponder why mobile platforms are
interesting from the standpoint of parallel and distributed computing,
even if at first glance it is obvious. From an architectural point of
view, the landscape of mobile devices has followed a similar
trajectory to that of traditional multiprocessing systems. The early
mobile device offerings, especially when it came to smartphones, were
single core. At the time of writing, the typical smartphone or tablet
is equipped with four CPU cores and a graphics processing unit (GPU),
with the trend of increasing cores (to at least 8) expected to continue in mobile CPUs. In
this vein, today's--and tomorrow's--devices need to be considered serious parallel systems in their own right. (In
fact, in the embedded space, there has been a corresponding emergence
of parallel boards, similar to the Raspberry Pi.)

The state of parallel computing today largely requires the mastery of
two styles, often appearing in a hybrid form: \emph{task parallelism}
and \emph{data parallelism}. The emerging mobile devices are following
desktop and server architecture by supporting both of these. In the
case of task parallelism, to get good performance, especially when it
comes to the user experience, concurrency must be disciplined. An
additional constraint placed on mobile devices, compared to parallel
computing, is that unbounded concurrency (threading) makes the device
unusable/unresponsive, even to a greater extent than on desktops and
servers (where there is better I/O performance in general). We posit
that learning to program concurrency in a resource-constrained
environment (e.g. Android smartphones) can be greatly helpful for
writing better concurrent, parallel, and distributed code in
general. More importantly, today's students really want to learn about
emerging platforms, so this is a great way to develop new talent in
languages and systems that are likely to be used in future
parallel/distributed programming environments.

\section{Roadmap}

In the remainder of this article, we first summarize the fundamentals
of thread safety in terms of concurrent access to shared mutable
state.  

We then discuss the technical domain of applications with graphical
user interfaces (GUIs), GUI application frameworks that target this
domain, and the runtime environment these frameworks typically
provide. 

Next, we examine a simple interactive behavior and explore how to
implement this using the Android mobile application devel	opment
framework. To the end of making our presentation relevant to problem
solvers, our running example is a bounded click counter application
(much more interactive and exciting than the examples commonly found
in concurrency textbooks, e.g., atomic counters, bounded buffers,
dining philosophers) that can be used to keep track of the capacity
of, say, a movie theater.

We then explore more interesting scenarios by introducing timers and
internal events. For example, a countdown timer can be used for
notification of elapsed time, a concept that has almost uniquely
emerged in the mobile space but has applications in embedded and
parallel computing in general, where asynchronous paradigms have been
present for some time, dating to job scheduling, especially for
longer-running jobs.

We close by exploring applications requiring longer-running, processor-bound activities. 
In mobile app development, a crucial design goal is to ensure UI responsiveness and appropriate progress reporting. 
We demonstrate techniques for ensuring computation proceeds but can be interrupted by the user. 
These techniques can be generalized to offload processor-bound activities to cloud-hosted web services.%
\footnote{This topic goes beyond the scope of this article but is included in the corresponding example~\cite{CS313PrimeChecker}.}

\section{Fundamentals of Thread Safety}
\label{sec:ThreadSafety}

Before we discuss concurrency issues in GUI applications, it is
helpful to understand the underlying fundamentals of thread safety in
situations where two or more concurrent threads (or other types of
activities) access shared mutable state.

Thread safety is best understood in terms of \emph{correctness}: An
implementation is correct if and only if it conforms to its
specification. The implementation is thread-safe if and only if it
continues to behave correctly in the presence of multiple
threads~\cite{JCIP}. 

\subsection{Example: incrementing a shared variable}

Let’s illustrate these concepts with perhaps the simplest possible
example: incrementing an integer number. The specification for this
behavior follows from the definition of increment: \emph{After performing
the increment, the number should be one greater than before}.

Here is a first attempt to implement this specification in the form of
an instance variable in a Java class and a \verb+Runnable+ instance that wraps
around our increment code and performs it on demand when we invoke its
run method (see below).
\begin{lstlisting}
int shared = 0;

final Runnable incrementUnsafe = new Runnable() {
  @Override
  public void run() {
    final int local = shared;
    tinyDelay();
    shared = local + 1;
  }
};
\end{lstlisting}

To test whether our implementation satisfies the specification, we can
write a simple test case:
\begin{lstlisting}
final int oldValue = shared;
incrementUnsafe.run();
assertEquals(oldValue + 1, shared);
\end{lstlisting}

In this test, we perform the increment operation in the only thread we
have, that is, the main thread. Our implementation passes the test
with flying colors. Does this mean it is thread-safe, though? 

To find out, we will now test two or more concurrent increment
operations, where the instance variable shared becomes shared
state. Generalizing from our specification, the value of the variable
should go up by one for each increment we perform. We can write this
test for two concurrent increments
\begin{lstlisting}
final int threadCount = 2;
final int oldValue = shared;
runConcurrently(incrementUnsafe, threadCount);
assertEquals(oldValue + threadCount, shared);
\end{lstlisting}
where \texttt{runConcurrently} runs the given code concurrently in the
desired number of threads: 
\begin{lstlisting}
public void runConcurrently(final Runnable inc, final int threadCount) {
  final Thread[] threads = new Thread[threadCount];
  for (int i = 0; i < threadCount; i += 1) {
    threads[i] = new Thread(inc);
  }
  for (final Thread t : threads) {
    t.start();
  }
  for (final Thread t : threads) {
    try {
      t.join();
    } catch (final InterruptedException e) {
      throw new RuntimeException("interrupted during join");
    }
  }
}
\end{lstlisting}
But this test does not always pass! When it does not, one of the two
increments appears to be lost. Even if its failure rate were one in a
million, the specification is violated, meaning that \emph{our
implementation of increment is not thread-safe}. 

\subsection{Interleaved versus serialized execution}

Let’s try to understand exactly what is going on here. We are
essentially running two concurrent instances of this code:
\begin{lstlisting}
/*f1*/ final int local1 = shared;    /*f2*/ final int local2 = shared;
/*s1*/ shared = local1 + 1;          /*s2*/ shared = local2 + 1;
\end{lstlisting}
(For clarity, we omit the invocation of \verb+tinyDelay+ present in the code
above; this invokes \verb+Thread.sleep(0)+ and is there just so we can
observe and discuss this phenomenon in conjunction with the Java
thread scheduler.)

The instructions are labeled $f_n$ and $s_n$ for \verb+fetch+ and \verb+set+,
respectively. Within each thread, execution proceeds sequentially, so
we are guaranteed that $f_1$ always comes before $s_1$ and $f_2$ always comes
before $s_2$. But we don’t have any guarantees about the relative order
across the two threads, so all of the following interleavings are
possible:
\begin{itemize}
\item \textbf{$f_1$ $s_1$} $f_2$ $s_2$: increments \verb+shared+ by 2
\item $f_1$ $f_2$ $s_1$ $s_2$: increments \verb+shared+ by 1
\item $f_1$ $f_2$ $s_2$ $s_1$: increments \verb+shared+ by 1
\item $f_2$ $f_1$ $s_1$ $s_2$: increments \verb+shared+ by 1
\item $f_2$ $f_1$ $s_2$ $s_1$: increments \verb+shared+ by 1
\item $f_2$ $s_2$ \textbf{$f_1$ $s_1$}: increments \verb+shared+ by 2
\end{itemize}
This kind of situation, where the behavior is nondeterministic in the
presence of two or more threads is also called a \emph{race condition}.\footnote{
When analyzing race conditions, we might be tempted to enumerate the
different possible interleavings. While it seems reasonable for our
example, this quickly becomes impractical because of the combinatorial
explosion for larger number of threads with more steps.} 

Based on our specification, the only correct result for incrementing
twice is to see the effect of the two increments, meaning the value of
shared goes up by two. Upon inspection of the possible interleavings
and their results, the only correct ones are those where both steps of
one increment happen before both steps of the other increment. 

Therefore, to make our implementation thread-safe, we need to make
sure that the two increments do not overlap. Each has to take place
\emph{atomically}. This requires one to go first and the other to go second;
their execution has to be \emph{serialized} or \emph{sequentialized} (see
also~\cite{AndroidMemVis} for details on the \emph{happens-before} relation
among operations on shared memory).


\subsection{Using locks to guarantee serialization}

In thread-based concurrent programming, the primary means to ensure
atomicity is mutual exclusion by \emph{locking}. Most thread implementations,
including \emph{p-threads (POSIX threads)}, provide some type of locking
mechanism.  

Because Java supports threads in the language, each object carries its
own lock, and there is a \verb+synchronized+ construct for allowing a thread
to execute a block of code only with the lock held. While one thread
holds the lock, other threads wanting to acquire the lock on the same
object will join the \emph{wait set} for that object. As soon as the lock
becomes available---when the thread currently holding the lock
finishes the synchronized block---, another thread from the wait set
receives the lock and proceeds. (In particular, there is no 
first-come-first-serve or other fairness guarantee for this wait set.)

We can use locking to make our implementation of increment atomic and
thereby thread-safe~\cite{CS313Threads}:
\begin{lstlisting}
final Object lock = new Object();

final Runnable incrementSafe = new Runnable() {
  @Override
  public void run() {
    synchronized (lock) {
      final int local = shared;
      tinyDelay();
      shared = local + 1;
    }
  }
};
\end{lstlisting}
Now it is guaranteed to pass the test every time.
\begin{lstlisting}
final int threadCount = 2;
final int oldValue = shared;
runConcurrently(incrementUnsafe, threadCount);
assertEquals(oldValue + threadCount, shared);
\end{lstlisting}
We should note that thread safety comes at a price: There is a small
but not insignificant overhead in handling locks and managing their
wait sets. 

\section{The GUI Programming Model and Runtime Environment}

As we mentioned above, common to most GUI application framework is the
need to balance ease of programming, correctness, performance, and
consistency of look-and-feel. In this section, we will discuss the
programming model and runtime environment of a typical GUI framework. 

In a GUI application, the user communicates with the application
through input events, such as button presses, menu item selections,
etc. The application responds to user events by invoking some piece of
code called an \emph{event handler} or \emph{event listener}. To send output back to
the user, the event handler typically performs some action that the
user can observe, e.g., displaying some text on the screen or playing
a sound.

\subsection{The GUI runtime environment}

\begin{figure}
\includegraphics[width=\textwidth]{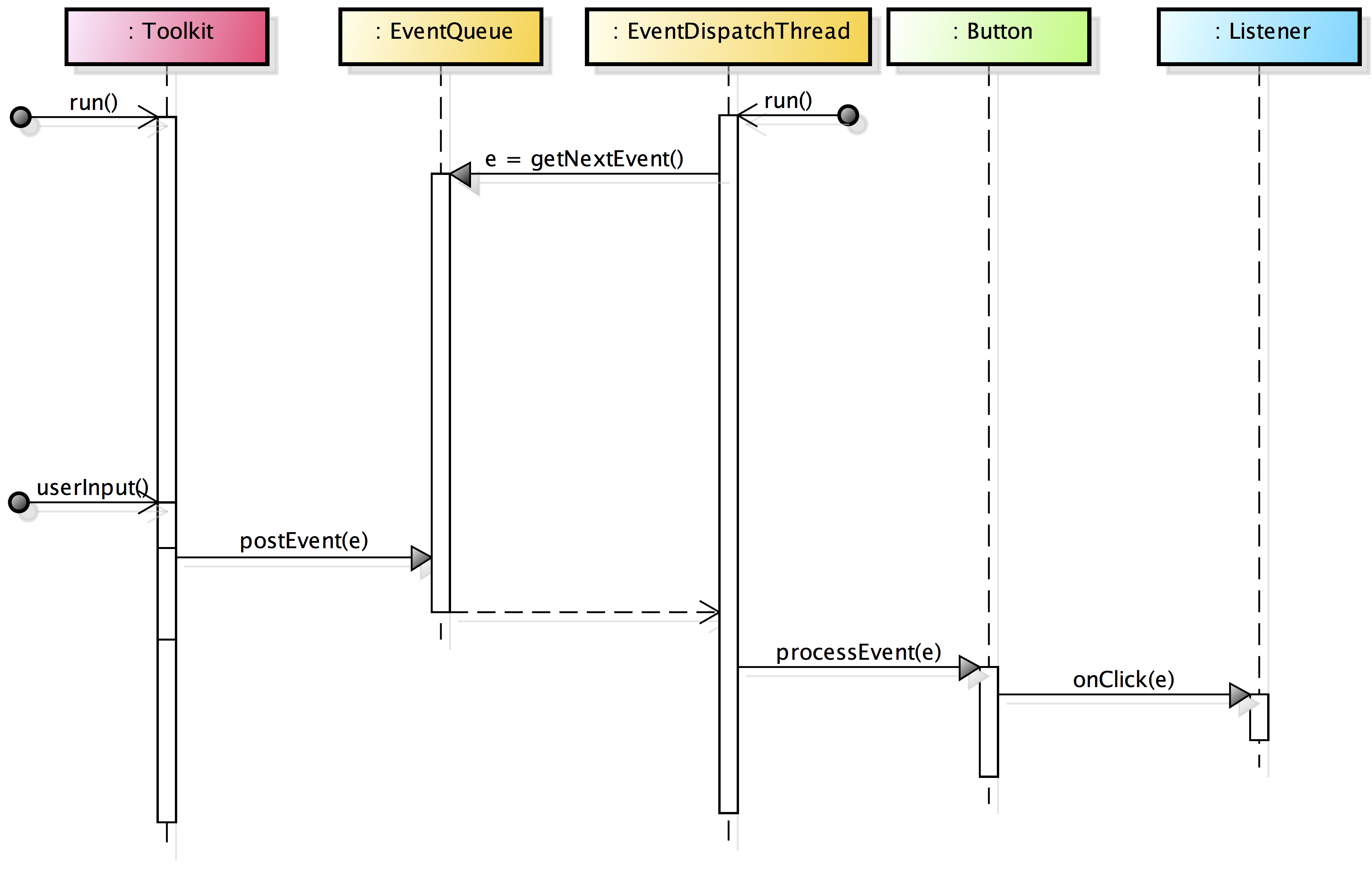}
\caption{UML sequence diagram showing the producer-consumer
architecture of a GUI. Stick arrowheads represent asynchronous
invocation, while solid arrowheads represent (synchronous) method
invocation.}
\label{fig:GUIArchitecture}
\end{figure}

Real-world GUI applications can be quite complex in terms of the
number of components and their logical containment hierarchy. The GUI
framework is responsible for translating \emph{low-level events} such as
mouse clicks and key presses to \emph{semantic events} such as button presses
and menu item selections targeting the correct component instances. To
manage this complexity, typical GUI frameworks use a producer-consumer
architecture, in which an internal, high-priority system thread places
low-level events on an event queue, while an application-facing 
\emph{UI thread}\footnote{In some frameworks, including Java AWT/Swing,
the UI thread is known as \emph{event dispatch thread (EDT)}.}
takes successive events from this queue and
delivers each event to its correct target component, which then
forward it to any attached listener(s). The UML sequence diagram
in figure~\ref{fig:GUIArchitecture} illustrates this architecture.

The event queue is a fully synchronized object. So it can be shared
safely between producer and consumer. It coalesces and filters groups
of events as appropriate, maintaining the following discipline: 
\begin{itemize}
\item \emph{Sequential (single-threaded)} processing: At most one event from
  this queue is dispatched simultaneously. 
\item \emph{Preservation of ordering:} If an event A is enqueued to the event
  queue before event B, then event B will not be dispatched before
  event A. 
\end{itemize}
Concretely, the UI thread continually takes events from
the event queue and processes them. Here is the pseudo-code for a
typical UI thread. 
\begin{lstlisting}
run() {
  while (true) {
    final Event event = eq.getNextEvent();
    final Object src = event.getSource();
    ((Component) src).processEvent(event);
  }
}
\end{lstlisting}
The target component, e.g., \verb+Button+, forwards events to its
listener(s).
\begin{lstlisting}
processEvent(e) {
  if (e instanceof OnClickEvent) {
    listener.onClick(e);
  }
  ...
}
\end{lstlisting}
While this presentation is mostly based on Java’s AWT for simplicity,
Android follows a similar approach with \verb+MessageQueue+ at the core and
some responsibilities split between \verb+Handler+ and \verb+Looper+
instances~\cite{EventQueue}.

This general approach, where requests (the events) come in
concurrently, get placed on a request queue, and are dispatched
sequentially to handlers, is an instance of the \emph{Reactor design
pattern}~\cite{POSA2}. 

\subsection{The application programmer’s perspective}

Within the GUI programming model, the application programmer focuses
on creating components and attaching event listeners to them. The
following is a very simple example of the round-trip flow of
information between the user and the application.
\begin{lstlisting}
final Button button = new Button("press me");
final TextView display = new TextView("hello");

increment.setOnClickListener(new OnClickListener() {
  public void onClick(final View view) {
    display.setText("world");
  }
});
\end{lstlisting}
The event listener design pattern at work here is an instance of the
\emph{Observer design pattern}~\cite{GOF}: Whenever the event source, such as the
button, has something to say, it notifies its observer(s) by invoking
the corresponding event handling method and passing itself as the
argument to this method. If desired, the listener can then obtain
additional information from the event source.

\subsection{Thread safety in GUI applications: the single-threaded
  rule}
\label{sec:SingleThreaded}

Generally, the programmer is oblivious to the concurrency between the
internal event producer thread and the EDT. The question is whether
there is or should be any concurrency on the application side. For
example, if two button presses occur in very short succession, can the
two resulting invocations of \verb+display.setText+ overlap in time and give
rise to thread safety concerns? In that case, shouldn’t we make the
GUI thread-safe by using locking? 

The answer is that typical GUI frameworks are already designed to
address this concern. Because a typical event listener accesses and/or
modifies the data structure constituting the visible GUI, if there
were concurrency among event listener invocations, we would have to
achieve thread safety by serializing access to the GUI using a lock
(and paying the price for this). It would be the application
programmer’s responsibility to use locking whenever an event listener
accesses the GUI. So we would have greatly complicated the whole model
without achieving significantly greater concurrency in our system.

We recall our underlying producer-consumer architecture, in which the
EDT processes one event at a time in its main loop. This means that
event listener invocations are already serialized. Therefore, we can
achieve thread safety directly and without placing an additional
burden on the programmer by adopting this simple rule: 

\begin{quote}
\emph{The application must always access GUI components from the UI thread. }
\end{quote}

This rule, known as the \emph{single-threaded rule}, is common among most GUI
frameworks, including Java Swing and Android. In practice, such access
must happen either during initialization (before the application
becomes visible), or within event listener code. Because it sometimes
becomes necessary to create additional threads (usually for
performance reasons), there are ways for those threads to schedule
code for execution on the EDT.  

Android actually \emph{enforces} the single-threaded GUI component access
rule by raising an exception if this rule is violated at
runtime. Android also enforces the ``opposite'' rule: It prohibits any
code on the event handling thread that will block it, such as network
access or database queries~\cite{AndroidThreads}.

\section{Single-Threaded Event-Based Applications}

In this section, we will study a large class class of applications
that won’t need any explicit concurrency at all. As long as each
response to an input event is short, we can keep these applications
simple and responsive by staying within the Reactor pattern.

We’ll start with a simple interactive behavior and explore how to
implement this using the Android mobile application development
framework~\cite{Android}. Our running example will be a bounded click
counter application that can be used to keep track of the capacity of,
say, a movie theater. The complete code for this example is available
online~\cite{CS313ClickCounter}.

\subsection{The bounded counter abstraction}

A \emph{bounded counter}~\cite{OS}, the concept underlying this application,
is an integer counter that is guaranteed to stay between a
preconfigured minimum and maximum value. This is called the \emph{data
invariant} of the bounded counter.
\[
  min \leq counter \leq max
\]

We can represent this abstraction as a simple, passive object with,
say, the following interface: 
\begin{lstlisting}
public interface BoundedCounter {
  void increment();
  void decrement();
  int get();
  boolean isFull();
  boolean isEmpty();
}
\end{lstlisting}

In following a \emph{test-driven} mindset~\cite{TDD}, we test
implementations of this interface using methods such as this one,
which ensures that incrementing the counter works properly: 
\begin{lstlisting}
@Test
public void testIncrement() {
  decrementIfFull();
  assertFalse(counter.isFull());
  final int v = counter.get();
  counter.increment();
  assertEquals(v + 1, counter.get());
}
\end{lstlisting}

In the remainder of this section, we'll put this abstraction to good
use by building an interactive application on top of it.

\subsection{The functional requirements for click counter device}

Next, let's imagine a device that realizes this bounded counter
concept. For example, a greeter positioned at the door of a movie
theater to prevent overcrowding, would require a device with the
following behavior:

\begin{itemize}
\item The device is preconfigured to the capacity of the venue.   
\item The device always displays the current counter value, initially
  zero.   
\item Whenever a person enters the movie theater, the greeter
  presses the \emph{increment} button; if there is still capacity, the
  counter value goes up by one.  
\item Whenever a person leaves the
  theater, the greeter presses the \emph{decrement} button; the counter value
  goes down by one (but not below zero).  
\item If the maximum has been
  reached, the \emph{increment} button either becomes unavailable (or, as an
  alternative design choice, attempts to press it cause an
  error). This behavior continues until the counter value falls below
  the maximum again.  
\item There is a \emph{reset} button for resetting the
  counter value directly to zero.
\end{itemize}

\subsection{A simple graphical user interface (GUI) for a click
  counter}

We now provide greater detail on the user interface of this click counter
device. In the case of a dedicated hardware device, the interface
could have tactile inputs and visual outputs, along with, say, audio
and haptic outputs. 

As a minimum, we require these interface elements:

\begin{itemize}
\item Three buttons, for incrementing and decrementing the counter
  value and for resetting it to zero.  
\item A numeric display of the
  current counter value.  Optionally, we would benefit from different
  types of feedback: 
\item Beep and/or vibrate when reaching the maximum
  counter value.  
\item Show the percentage of capacity as a numeric
  percentage or color thermometer.
\end{itemize}
Instead of a hardware device, we'll now implement this behavior as a
mobile software app, so let's focus first on the minimum interface
elements. In addition, we'll make the design choice to disable operations
that would violate the counter's data invariant.

\begin{figure}
\begin{center}
\subfigure[Minimum state]{\includegraphics[width=0.3\textwidth]{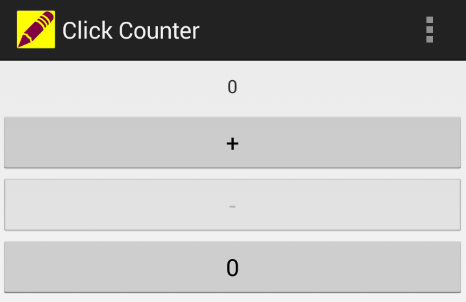}}
\subfigure[Counting state]{\includegraphics[width=0.3\textwidth]{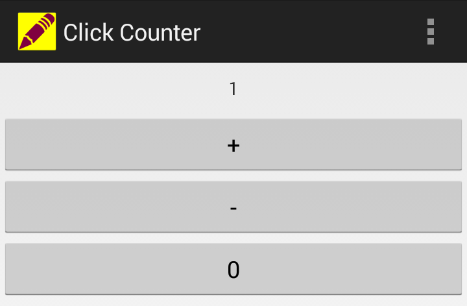}}
\subfigure[Maximum state]{\includegraphics[width=0.3\textwidth]{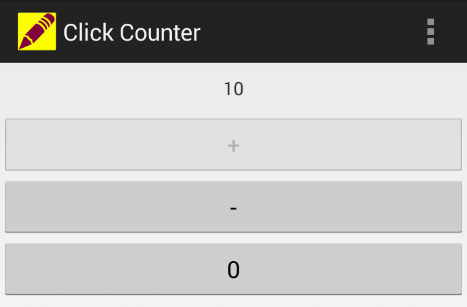}}
\end{center}
\caption{View states for the click counter}
\label{fig:BoundedCounterViewStates}
\end{figure}

These decisions lead to the three \emph{view states} for the bounded click
counter Android app (see figure~\ref{fig:BoundedCounterViewStates}: In
the initial (minimum) view state, the decrement button is disabled. In
the counting view state of the, all buttons are enabled. Finally, in
the maximum view state, the increment button is disabled; we assume a
maximum value of 10). In our design, the reset button is always
enabled.

\subsection{Understanding user interaction as events}
\label{sec:Events}

It was fairly easy to express the familiar bounded counter abstraction
and to envision a possible user interface for putting this abstraction
to practical use. The remaining challenge is to tie the two together
in a meaningful way, such that the interface uses the abstraction to
provide the required behavior. In this section, we'll work on bridging
this gap.

\subsubsection{Modeling the interactive behavior}

As a first step, let's abstract away the concrete aspects of the user
interface:
\begin{itemize}
\item Instead of touch buttons, we'll have \emph{input events}.  
\item Instead of setting a visual display, we'll \emph{modify a counter value}.
\end{itemize}
After we take this step, we can use a UML state machine
diagram~\cite{UML} to model the dynamic behavior we described at the
beginning of this section more formally.\footnote{
A full introduction to the Unified Modeling Language
(UML)~\cite{UML} would go far beyond the scope of this
article. Therefore, we aim to introduce the key elements of UML needed
here in an informal and pragmatic manner. Various UML resources,
including the official specification, are available at
\url{http://www.uml.org/}. Third-party tutorials are available online and in
book form.}
Note how the touch buttons
correspond to events (triggers of \emph{transitions}, i.e., arrows) with the
matching names.

\begin{figure}
\begin{center}
\includegraphics[width=\textwidth]{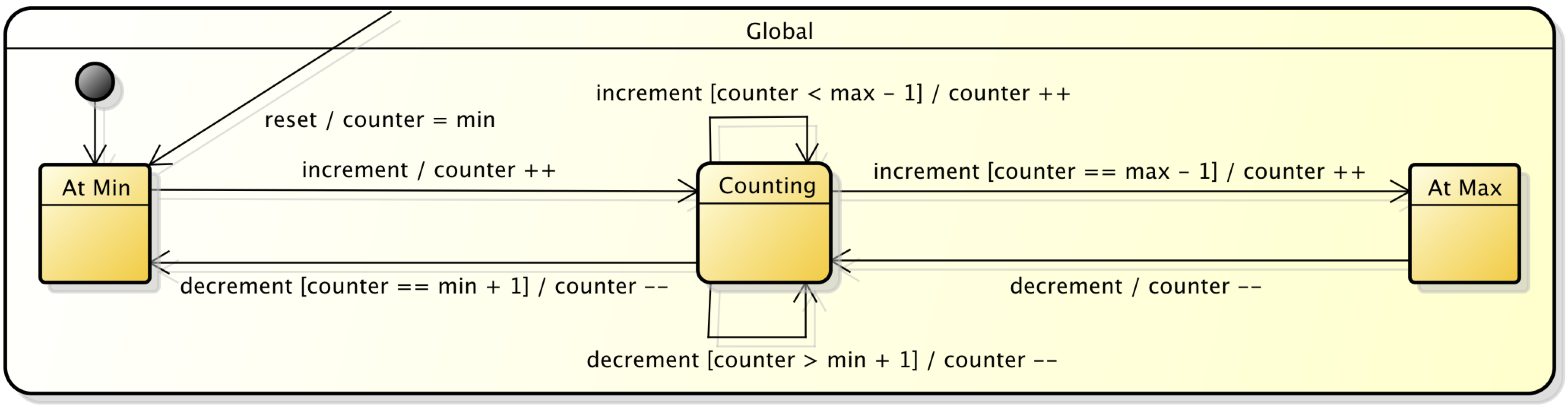}
\end{center}
\caption{UML state machine diagram modeling the dynamic behavior of
  the bounded counter application}
\label{fig:BoundedCounterStates}
\end{figure}

The behavior starts with the \emph{initial pseudostate} represented by the
black circle. From there, the counter value gets its initial value,
and we start in the minimum state. Assuming that the minimum and
maximum values are at least two apart, we can increment
unconditionally and reach the counting state. As we keep incrementing,
we stay here as long as we are at least two away from the maximum
state. As soon as we are exactly one away from the maximum state, the
next increment takes us to that state, and now we can no longer
increment, just decrement. The system mirrors this behavior in
response to the decrement event. There is a surrounding global state
to support a single reset transition back to the minimum state. 
Figure~\ref{fig:BoundedCounterStates} shows the complete diagram.

As you can see, the three model states map directly to the view states
from the previous subsection, and the transitions enabled in each
model state map to the buttons enabled in each view state. This is not
always the case, though, and we will see examples in a later section
of an application with multiple model states but only a single view
state.

\subsubsection{GUI components as event sources}


Our next step is to bring the app to life by connecting the visual
interface to the interactive behavior. For example, when pressing
the increment button in a non-full counter state, we expect the
displayed value to go up by one. In general, the user can trigger
certain events by interacting with view components and other event
sources. For example, one can press a button, swipe one's finger
across the screen, rotate the device, etc.

\subsubsection{Event listeners and the Observer pattern}

We now discuss what an event is and what happens after it gets
triggered. We will continue focusing on our running example of
pressing the increment button.

The visual representation of an Android GUI is usually auto-generated from an XML source during the build process.%
\footnote{It is also possible---though less practical---to build an Android GUI programatically.}
For example, the source element for our increment button looks
like this; it declaratively maps the \verb+onClick+ attribute to the \verb+onIncrement+
method in the associated activity instance.
\begin{lstlisting}[language=XML]
<Button
  android:id="@+id/button_increment"
  android:layout_width="fill_parent"
  android:layout_height="wrap_content"
  android:onClick="onIncrement"
  android:text="@string/label_increment" />
\end{lstlisting}
The \emph{Android manifest} associates an app with its main activity class. 
The top-level
\verb+manifest+ element specifies the Java package of the activity class, and
the activity element on line 5 specifies the name of the activity
class, \verb+ClickCounterActivity+. 
\begin{lstlisting}[language=XML]
<manifest xmlns:android="http://schemas.android.com/apk/res/android"
  package="edu.luc.etl.cs313.android.clickcounter" ...>
  ...
  <application ...>
    <activity android:name=".ClickCounterActivity" ...>
      <intent-filter>
        <action android:name="android.intent.action.MAIN" />
        <category android:name="android.intent.category.LAUNCHER" />
      </intent-filter>
    </activity>
  </application>
</manifest>
\end{lstlisting}
So an \emph{event} is just an invocation of an \emph{event listener} method,
possibly with an argument describing the event. We first need to
establish the association between an event source and one (or possibly
several) event listener(s) by \emph{subscribing} the listener to the
source. Once we do that, every time this source emits an event,
normally triggered by the user, the appropriate event listener method
gets called on each subscribed listener. 

Unlike ordinary method invocations, where the caller knows the
identity of the callee, the (observable) event source provides a
general mechanism for subscribing a listener to a source. This
technique is widely known as the \emph{Observer design pattern}~\cite{GOF}.

Many GUI frameworks follow this approach. In Android, for example, the
general component superclass is View, and there are various types of
listener interfaces, including \verb+OnClickListener+. In following the
\emph{Dependency Inversion Principle (DIP)}~\cite{APPP}, the \verb+View+ class owns
the interfaces its listeners must implement.

\begin{lstlisting}
public class View {
   ...
   public static interface OnClickListener {
       void onClick(View source);
   }
   public void setOnClickListener(final OnClickListener listener) { ... }
   ...
}
\end{lstlisting}
Android follows an event source/listener naming idiom loosely based on
the JavaBeans specification~\cite{JavaBeans}. Listeners of, say, the
\verb+onX+ event implement the \verb+OnXListener+ interface with the 
\verb+onX(Source source)+
method. Sources of this kind of event implement the
\verb+setOnXListener+ method.\footnote{
Readers who have worked with GUI framework that supports
multiple listeners, such as Swing, might initially find it restrictive
of Android to allow only one. We'll leave it as an exercise to figure
out which well-known software design pattern can be used to work
around this restriction.}
An actual event instance corresponds to an
invocation of the \verb+onX+ method with the source component passed as the
source argument.

\subsubsection{Processing events triggered by the user}

\begin{figure}
\begin{center}
\includegraphics[width=\textwidth]{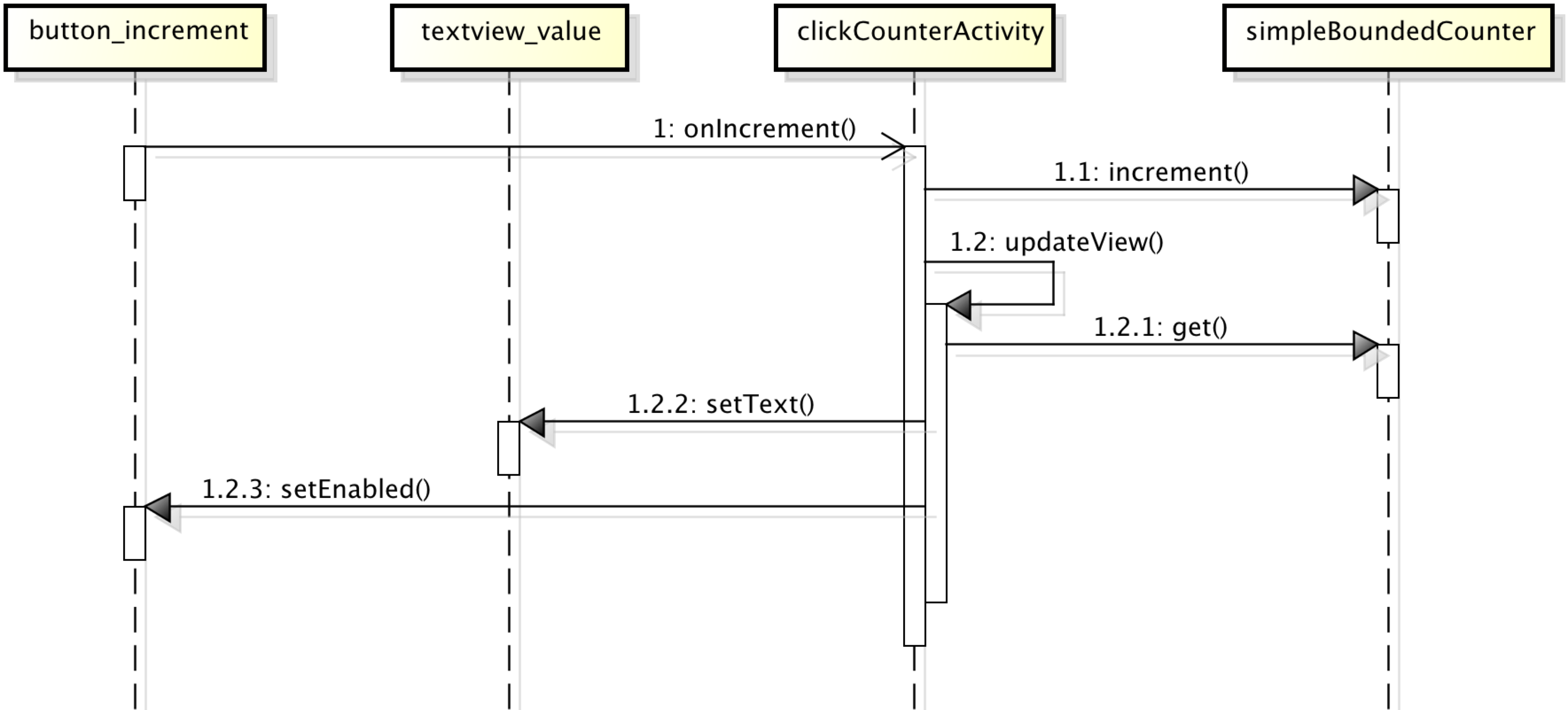}
\end{center}
\caption{Sequence diagram showing the full event-based interaction
cycle in response to a press of the increment button. Stick arrowheads
represent events, while solid arrowheads represent (synchronous)
method invocation.}
\label{fig:ClickCounterSequenceDiagram}
\end{figure}

The Android activity is responsible for mediating between the view
components and the POJO (plain old Java object) bounded counter model
we saw above. The full cycle of each event-based interaction goes like
this. By pressing the increment button, the user triggers the \verb+onClick+
event on that button, and the \verb+onIncrement+ method gets called. This
method interacts with the model instance by invoking the increment
method and then requests a view update of the activity itself. The
corresponding \verb+updateView+ method also interacts with the model instance
by retrieving the current counter value using the get method, displays
this value in the corresponding GUI element with unique ID
\verb+textview_value+, and finally updates the view states as necessary.
Figure~\ref{fig:ClickCounterSequenceDiagram} illustrates this
interaction step-by-step.
\begin{lstlisting}
public void onIncrement(final View view) {
  model.increment();
  updateView();
}
protected void updateView() {
  final TextView valueView = (TextView) findViewById(R.id.textview_value);
  valueView.setText(Integer.toString(model.get()));
  // afford controls according to model state
  ((Button) findViewById(R.id.button_increment)).setEnabled(!model.isFull());
  ((Button) findViewById(R.id.button_decrement)).setEnabled(!model.isEmpty());
}
\end{lstlisting}

\emph{What happens if the user presses two buttons at the same time?} As
discussed above, the GUI framework responds to at most one button
press or other event trigger at any given time. While the GUI
framework is processing an event, it places additional incoming event
triggers on a queue and fully processes each one in
turn. Specifically, only after the event listener method handling the
current event returns will the framework process the next
event. (Accordingly, activation boxes of different event listener
method invocations in the UML sequence diagram must not overlap.) This
approach is called \emph{single-threaded event handling}. It keeps the
programming model simple and avoids problems such as race conditions
or deadlocks that can arise in multithreaded approaches.

\subsection{Application architecture}

\begin{figure}
\begin{center}
\includegraphics[height=0.3\textwidth]{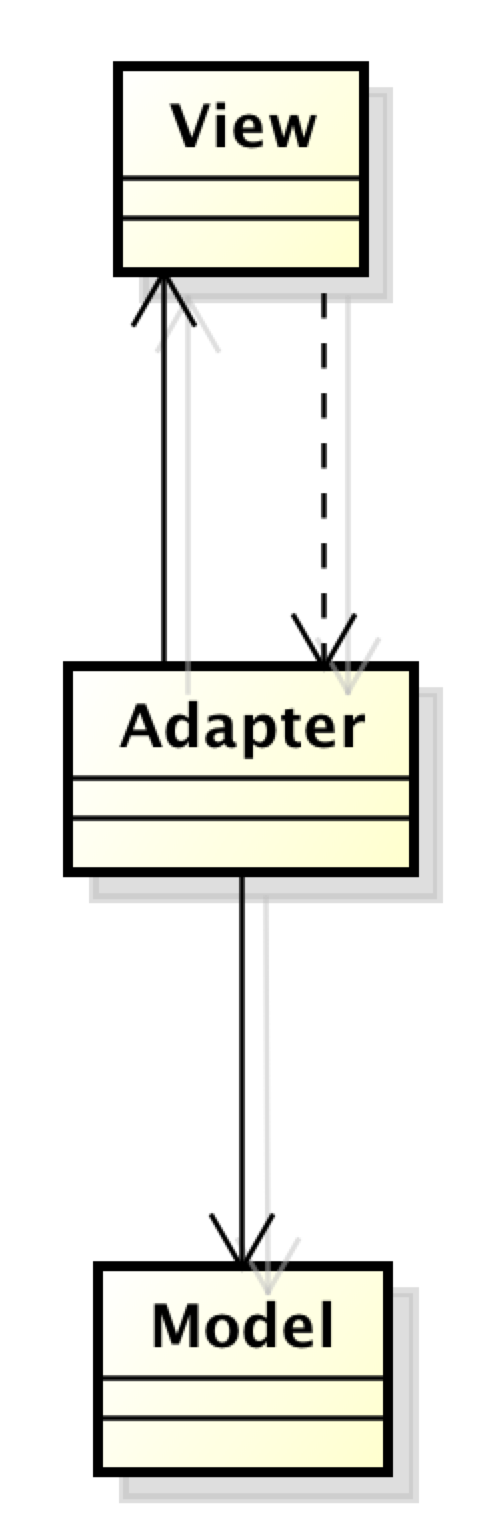}
\end{center}
\caption{UML class diagram showing the Model-View-Adapter (MVA)
architecture of the bounded click counter Android app. Solid arrows
represent method invocation, and dashed arrows represent event flow.}
\label{fig:ModelViewAdapter}
\end{figure}

This overall application architecture, where a component mediates
between view components and model components, is known as
\emph{Model-View-Adapter (MVA)}~\cite{MVA}, where the adapter component
mediates all interactions between the view and the model. (By
contrast, the \emph{Model-View-Controller (MVC)} architecture has a
triangular shape and allows the model to update the view(s) directly
via update events.) Figure~\ref{fig:ModelViewAdapter} illustrates this
architecture. The solid arrows represent ordinary method invocations,
and the dashed arrow represents event-based interaction. View and
adapter play the roles of observable and observer, respectively, in
the Observer pattern that describes the top half of this architecture.

\subsection{System-testing GUI applications}

Automated system testing of entire GUI applications is a broad and
important topic that goes beyond the scope of this article. Here, we
complete our running example by focusing on a few key concepts and
techniques. 


In system testing, we distinguish between our application code, usually referred to as
the \emph{system under test (SUT)}, and the \emph{test code}. 
At the beginning of this section, we already saw an example of a
simple component-level unit test method for the POJO bounded counter
model. Because Android view components support triggering events
programmatically, we can also write system-level test methods that
mimic the way a human user would interact with the application.

\subsubsection{System-testing the click counter}

The following test handles a simple scenario of pressing the reset
button, verifying that we are in the minimum view state, then pressing
the increment button, verifying that the value has gone up and we are
in the counting state, pressing the reset button again, and finally
verifying that we are back in the minimum state. 

\begin{lstlisting}
@Test
public void testActivityScenarioIncReset() {
  assertTrue(getResetButton().performClick());
  assertEquals(0, getDisplayedValue());
  assertTrue(getIncButton().isEnabled());
  assertFalse(getDecButton().isEnabled());
  assertTrue(getResetButton().isEnabled());
  assertTrue(getIncButton().performClick());
  assertEquals(1, getDisplayedValue());
  assertTrue(getIncButton().isEnabled());
  assertTrue(getDecButton().isEnabled());
  assertTrue(getResetButton().isEnabled());
  assertTrue(getResetButton().performClick());
  assertEquals(0, getDisplayedValue());
  assertTrue(getIncButton().isEnabled());
  assertFalse(getDecButton().isEnabled());
  assertTrue(getResetButton().isEnabled());
  assertTrue(getResetButton().performClick());
}
\end{lstlisting}
The next test ensures that the visible application state is preserved
under device rotation. This is an important and effective test because
an Android application goes through its entire lifecycle under
rotation.

\begin{lstlisting}
@Test
public void testActivityScenarioRotation() {
  assertTrue(getResetButton().performClick());
  assertEquals(0, getDisplayedValue());
  assertTrue(getIncButton().performClick());
  assertTrue(getIncButton().performClick());
  assertTrue(getIncButton().performClick());
  assertEquals(3, getDisplayedValue());
  getActivity().
    setRequestedOrientation(ActivityInfo.SCREEN_ORIENTATION_LANDSCAPE);
  assertEquals(3, getDisplayedValue());
  getActivity().
    setRequestedOrientation(ActivityInfo.SCREEN_ORIENTATION_PORTRAIT);
  assertEquals(3, getDisplayedValue());
  assertTrue(getResetButton().performClick());
}
\end{lstlisting}

\subsubsection{System testing in and out of container}

We have two main choices for system-testing our app:
\begin{itemize}
\item \emph{In-container/instrumentation testing} in the presence of the
  target execution environment, such as an actual Android phone or
  tablet emulator (or physical device). This requires deploying both
  the SUT and the test code to the emulator and tends to be quite
  slow. So far, Android's build tools officially support only this
  mode.  
\item \emph{Out-of-container testing} on the development workstation
  using a test framework such as \emph{Robolectric} that simulates an Android
  runtime environment tends to be considerably faster. This and other
  non-instrumentation types of testing can be integrated in the
  Android build process with a bit of extra effort.
\end{itemize}
Although the Android build process does not officially support this or
other types of non-instrumentation testing, they can be integrated in
the Android build process with a bit of extra effort. 

\subsubsection{Structuring test code for flexibility and reuse}

Typically, we'll want to run the exact same test logic in both cases,
starting with the simulated environment and occasionally targeting the
emulator or device. An effective way to structure our test code for
this purpose is the xUnit design pattern 
\emph{Testcase Superclass}~\cite{XUnitPatterns}. 
As the pattern
name suggests, we pull up the common test code into an abstract
superclass, and each of the two concrete test classes inherits the
common code and runs it in the desired environment.

\begin{lstlisting}
@RunWith(RobolectricTestRunner.class)
public class ClickCounterActivityRobolectric extends AbstractClickCounterActivityTest {
 // some minimal Robolectric-specific code
}
\end{lstlisting}
The official Android test support, however, requires inheriting from a
specific superclass called \verb+ActivityInstrumentationTestCase2+. This
class now takes up the only superclass slot, so we cannot use the
Testcase Superclass pattern literally. Instead, we need to approximate
inheriting from our \verb+AbstractClickCounterActivityTest+ using delegation
to a subobject. This gets the job done but can get quite tedious when
a lot of test methods are involved.

\begin{lstlisting}
public class ClickCounterActivityTest 
  extends ActivityInstrumentationTestCase2<ClickCounterActivity> {
  ...
  // test subclass instance to delegate to
  private AbstractClickCounterActivityTest actualTest;

  @UiThreadTest
  public void testActivityScenarioIncReset() {
    actualTest.testActivityScenarioIncReset();
  }
  ...
}
\end{lstlisting}
Having a modular architecture, such as model-view-adapter, enables us
to test most of the application components in isolation. For example,
our simple unit tests for the POJO bounded counter model still work in
the context of the overall Android app.

\subsubsection{Test coverage}
\emph{Test coverage} describes the extent to which our test code exercises the system under test, and there are several ways to measure test coverage~\cite{Zhu:1997:SUT:267580.267590}.
We generally want test coverage to be as close to 100\% as possible and can measure this using suitable tools, such as JaCoCo along with the corresponding Gradle plugin.%
\footnote{More information on JaCoCo the JaCoCo Gradle plugin is available at \url{http://www.eclemma.org/jacoco/} and \url{https://github.com/arturdm/jacoco-android-gradle-plugin}, respectively.}

\section{Interactive Behaviors and Implicit Concurrency with Internal
  Timers}

In this section, we'll study applications that have richer,
timer-based behaviors compared to the previous section. Our example
will be a countdown timer for cooking and similar scenarios where we
want to be notified when a set amount of time has elapsed. The
complete code for a very similar example is available
online~\cite{CS313Stopwatch}.

\subsection{The functional requirements for a countdown timer}

Let's start with the functional requirements for the countdown timer,
amounting to a fairly abstract description of its controls and
behavior.

The timer exposes the following controls:
\begin{itemize}
\item One two-digit display of the form 88.  
\item One multi-function button.  
\end{itemize}

The timer behaves as follows: 
\begin{itemize}
\item The timer always displays the remaining time in seconds.  
\item Initially, the timer is stopped and the (remaining) time is zero.  
\item If the button is pressed when the timer is stopped, the time is
  incremented by one up to a preset maximum of 99. (The button acts as
  an increment button.)   
\item If the  time is greater than zero and three seconds elapse from
  the most  recent time the button was pressed, then the timer beeps
  once and  starts running.   
\item While running, the timer subtracts one from the time for every
  second that elapses.   
\item If the timer is running and
  the button is pressed, the timer stops and the time is reset to
  zero. (The button acts as a cancel button.)  
\item If the timer is
  running and the time reaches zero by itself (without the button
  being pressed), then the timer stops counting down, and the alarm
  starts beeping continually and indefinitely.  
\item If the alarm is
  sounding and the button is pressed, the alarm stops sounding; the
  timer is now stopped and the (remaining) time is zero. (The button
  acts as a stop button.)
\end{itemize}

\subsection{A graphical user interface (GUI) for a countdown timer}

Our next step is to flesh out the GUI for our timer. For usability,
we'll label the multifunction button with its current function. We'll
also indicate which state the timer is currently in.

\begin{figure}
\begin{center}
\subfigure[initial stopped state with zero time]{\includegraphics[height=0.25\textwidth]{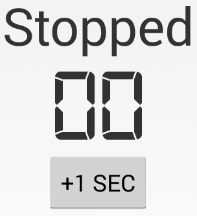}}
\subfigure[initial stopped state after adding some time]{\includegraphics[height=0.25\textwidth]{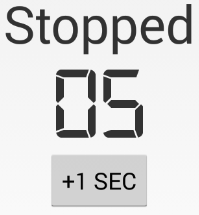}}
\subfigure[running (counting down) state]{\includegraphics[height=0.25\textwidth]{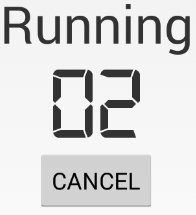}}
\subfigure[alarm ringing state]{\includegraphics[height=0.25\textwidth]{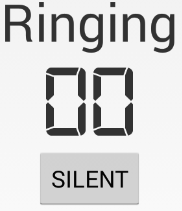}}
\end{center}
\caption{View states for the countdown timer}
\label{fig:CountdownTimerViewStates}
\end{figure}

The screenshots in figure~\ref{fig:CountdownTimerViewStates} show the
default scenario where we start up the timer, add a few seconds, wait
for it to start counting down, and ultimately reach the alarm state.

\subsection{Modeling the interactive behavior}

Let's again try to describe the abstract behavior of the countdown
timer using a UML state machine diagram. As usual, there are various
ways to do this, and our guiding principle is to keep things simple
and close to the informal description of the behavior.

\begin{figure}
\includegraphics[width=\textwidth]{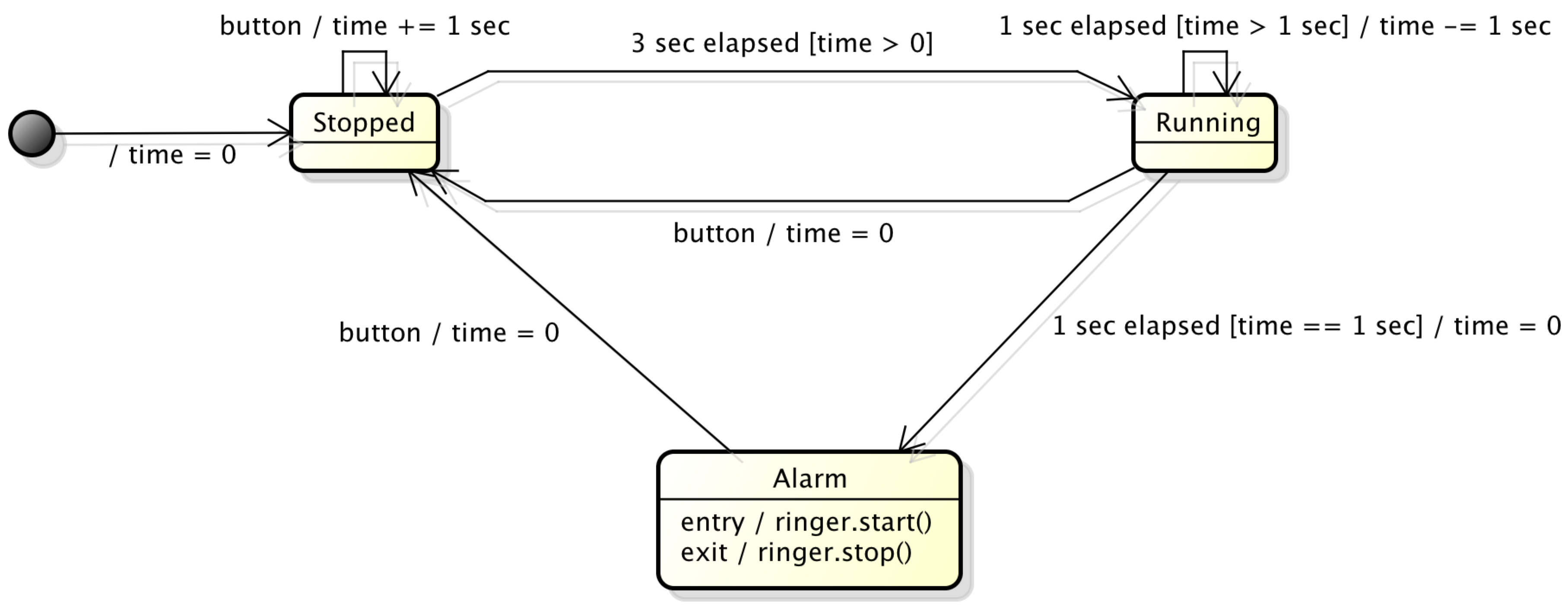}
\caption{UML state machine diagram modeling the dynamic behavior
of the countdown timer application}
\label{fig:CountdownTimerStates}
\end{figure}

It is easy to see that we need to represent the current counter
value. Once we accept this, we really don't need to distinguish
between the stopped state (with counter value zero) and the counting
state (with counter value greater than zero). The other states that
arise naturally are the running state and the alarm
state. Figure~\ref{fig:CountdownTimerStates} shows the resulting UML
state machine diagram.

As in the click counter example, these model states map directly to
the view states shown above. Again, the differences among the view
states are very minor and are aimed mostly at usability: A properly
labeled button is a much more effective affordance than an unlabeled
or generically labeled one.

Note that there are two types of (internal) timers at work here:
\begin{itemize}
\item \emph{one-shot timers}, such as the three-second timer in the stopped state that gets restarted every time we press the multifunction button to add time
\item \emph{recurring timers}, such as the one-second timer in the running
  state that fires continually for every second that goes by
\end{itemize}

The following is the control method that starts a recurring timer that
ticks approximately every second.

\begin{lstlisting}
// called on the UI thread
public void startTick(final int periodInSec) {
  if (recurring != null) throw new IllegalStateException();

  recurring = new Timer();

  // The clock model runs onTick every 1000 milliseconds
  recurring.schedule(new TimerTask() {
    @Override
    public void run() {
      // fire event on the timer’s internal thread
      listener.onTick();
    }
  }, periodInSec * 1000, periodInSec * 1000); // initial and periodic delays
}
\end{lstlisting}

\subsection{Thread-safety in the model}

Within the application model, each timer has its own internal thread
on which it schedules the run method of its \verb+TimerTask+
instances. Therefore, other model components, such as the state
machine, that receive events from either the UI and one or more
timers, or more than one timer, will have to be kept thread-safe. The
easiest way to achieve this is to use locking by making all relevant
methods in the state machine object synchronized; this design pattern
is known as\emph{Fully Synchronized Object}~\cite{CPJ2E} or 
\emph{Monitor Object}~\cite{HPJPC,POSA2,JCIP}. 
\begin{lstlisting}
@Override public synchronized void onButtonPress() { state.onButtonPress(); }
@Override public synchronized void onTick()        { state.onTick(); }
@Override public synchronized void onTimeout()     { state.onTimeout(); }
\end{lstlisting}
Furthermore, update events coming back into the adapter component of
the UI may happen on one of the timer threads. Therefore, to comply
with the single-threaded rule, the adapter has to explicitly
reschedule such events on the UI thread, using the \verb+runOnUiThread+
method it inherits from \verb+android.app.Activity+.

\begin{lstlisting}
@Override
public void updateTime(final int time) {
  // UI adapter responsibility to schedule incoming events on UI thread
  runOnUiThread(new Runnable() {
    @Override
    public void run() {
      final TextView tvS = (TextView) findViewById(R.id.seconds);
      tvS.setText(Integer.toString(time / 10) + Integer.toString(time % 10));
    }
  });
}
\end{lstlisting}

Alternatively, you may wonder whether we can stay true to the
single-threaded rule and reschedule all events on the UI thread at
their sources. This is possible using mechanisms such as the
\verb+runOnUiThread+ method and has the advantage that the other model
components such as the state machine no longer have to be
thread-safe. The event sources, however, would now depend on the
adapter; to keep this dependency manageable and our event sources
testable, we can express it in terms of a small interface (to be
implemented by the adapter) and inject it into the event sources. 
\begin{lstlisting}
public interface UIThreadScheduler {
   void runOnUiThread(Runnable r);
}
\end{lstlisting}

Some GUI frameworks, such as Java Swing, provide non-view components
for scheduling tasks or events on the UI thread, such as
\verb+javax.swing.Timer+. This avoids the need for an explicit dependency on
the adapter but retains the implicit dependency on the UI layer.

Meanwhile, Android developers are being encouraged to use
\verb+ScheduledThreadPoolExecutor+ instead of \verb+java.util.Timer+, though the
thread-safety concerns remain the same as before.

\subsection{Implementing time-based autonomous behavior}

While the entirely passive bounded counter behavior from the previous
section was straightforward to implement, the countdown timer includes
autonomous timer-based behaviors that give rise to another level of
complexity.

There are different ways to deal with this behavioral
complexity. Given that we have already expressed the behavior as a
state machine, we can use the \emph{State design pattern}~\cite{GOF} to
separate state-dependent behavior from overarching handling of
external and internal triggers and actions.

We start by defining a state abstraction. Besides the same common
methods and reference to its surrounding state machine, each state has
a unique identifier.

\begin{lstlisting}
abstract class TimerState implements TimerUIListener, ClockListener {

  public TimerState(final TimerStateMachine sm) { this.sm = sm; }

  protected final TimerStateMachine sm;

  @Override public final void onStart() { onEntry(); }    
  public void onEntry() { }
  public void onExit() { }
  public void onButtonPress() { }
  public void onTick() { }
  public void onTimeout() { }
  public abstract int getId();
}
\end{lstlisting}
In addition, a state receives UI events and clock ticks. Accordingly, it implements the corresponding interfaces, which are defined as follows:
\begin{lstlisting}
public interface TimerUIListener {
  void onStart();
  void onButtonPress();
}

public interface ClockListener { 
  void onTick();
  void onTimeout(); 
}
\end{lstlisting}
As we discussed in section~\ref{sec:Events}, Android follows an event
source/listener naming idiom. As our examples illustrate, it is
straightforward to define custom app-specific events that follow this
same convention. Our \verb+ClockListener+, for example, combines two kinds of
events within a single interface.

Concrete state classes implement the abstract \verb+TimerState+ class. The
key parts of the state machine implementation follow: 
\begin{lstlisting}
private TimerState state = new TimerState(this) { // intial pseudo-state 
   @Override public int getId() { throw new IllegalStateException(); }
};

protected void setState(final TimerState nextState) { 
   state.onExit();
   state = nextState;
   uiUpdateListener.updateState(state.getId());
   state.onEntry();
}
\end{lstlisting}

Let's focus on the stopped state first. In this state, neither is the
clock ticking, nor is the alarm ringing. On every button press, the
remaining running time goes up by one second and the one-shot
three-second idle timeout starts from zero. If three seconds elapse
before another button press, we transition to the running state.
\begin{lstlisting}
private final TimerState STOPPED = new TimerState(this) {
  @Override public void onEntry()   { timeModel.reset(); updateUIRuntime(); } 
  @Override public void onButtonPress() {
      clockModel.restartTimeout(3 /* seconds */);
      timeModel.inc(); updateUIRuntime();
  }
  @Override public void onTimeout() { setState(RUNNING); }
  @Override public int getId()      { return R.string.STOPPED; } 
};
\end{lstlisting}

Let's now take a look at the running state. In this state, the clock
is ticking but the alarm is not ringing. With every recurring clock
tick, the remaining running time goes down by one second. If it
reaches zero, we transition to the ringing state. If a button press
occurs, we stop the clock and transition to the stopped state.
\begin{lstlisting}
private final TimerState RUNNING = new TimerState(this) {
  @Override public void onEntry()       { clockModel.startTick(1 /* second */); }
  @Override public void onExit()        { clockModel.stopTick(); }
  @Override public void onButtonPress() { setState(STOPPED); }
  @Override public void onTick() {
      timeModel.dec(); updateUIRuntime();
      if (timeModel.get() == 0) { setState(RINGING); } 
  }
  @Override public int getId()          { return R.string.RUNNING; } 
};
\end{lstlisting}

Finally, in the ringing state, nothing is happening other than the
alarm ringing. If a button press occurs, we stop the alarm and
transition to the stopped state.
\begin{lstlisting}
private final TimerState RINGING = new TimerState(this) {
  @Override public void onEntry()       { uiUpdateListener.ringAlarm(true); } 
  @Override public void onExit()        { uiUpdateListener.ringAlarm(false); } 
  @Override public void onButtonPress() { setState(STOPPED); }
  @Override public int getId()          { return R.string.RINGING; }
};
\end{lstlisting}

\subsection{Managing structural complexity}

\begin{figure}
\begin{center}
\includegraphics[height=0.3\textwidth]{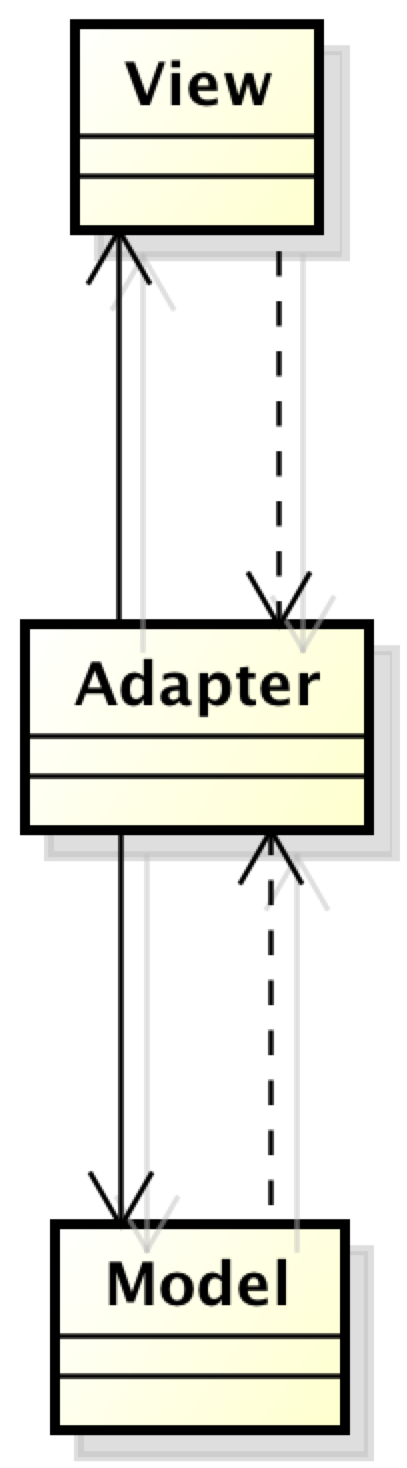}
\end{center}
\caption{The countdown timer’s Model-View-Adapter (MVA) architecture with additional event flow from model to view}
\label{fig:AutonomousModelViewAdapter}
\end{figure}

We can again describe the architecture of the countdown timer Android
app as an instance of the Model-View-Adapter (MVA) architectural
pattern. In figure~\ref{fig:AutonomousModelViewAdapter}, solid arrows
represent (synchronous) method invocation, and dashed arrows represent
(asynchronous) events. Here, both the view components and the model's
autonomous timer send events to the adapter.

\begin{figure}
\begin{center}
\includegraphics[height=0.4\textheight]{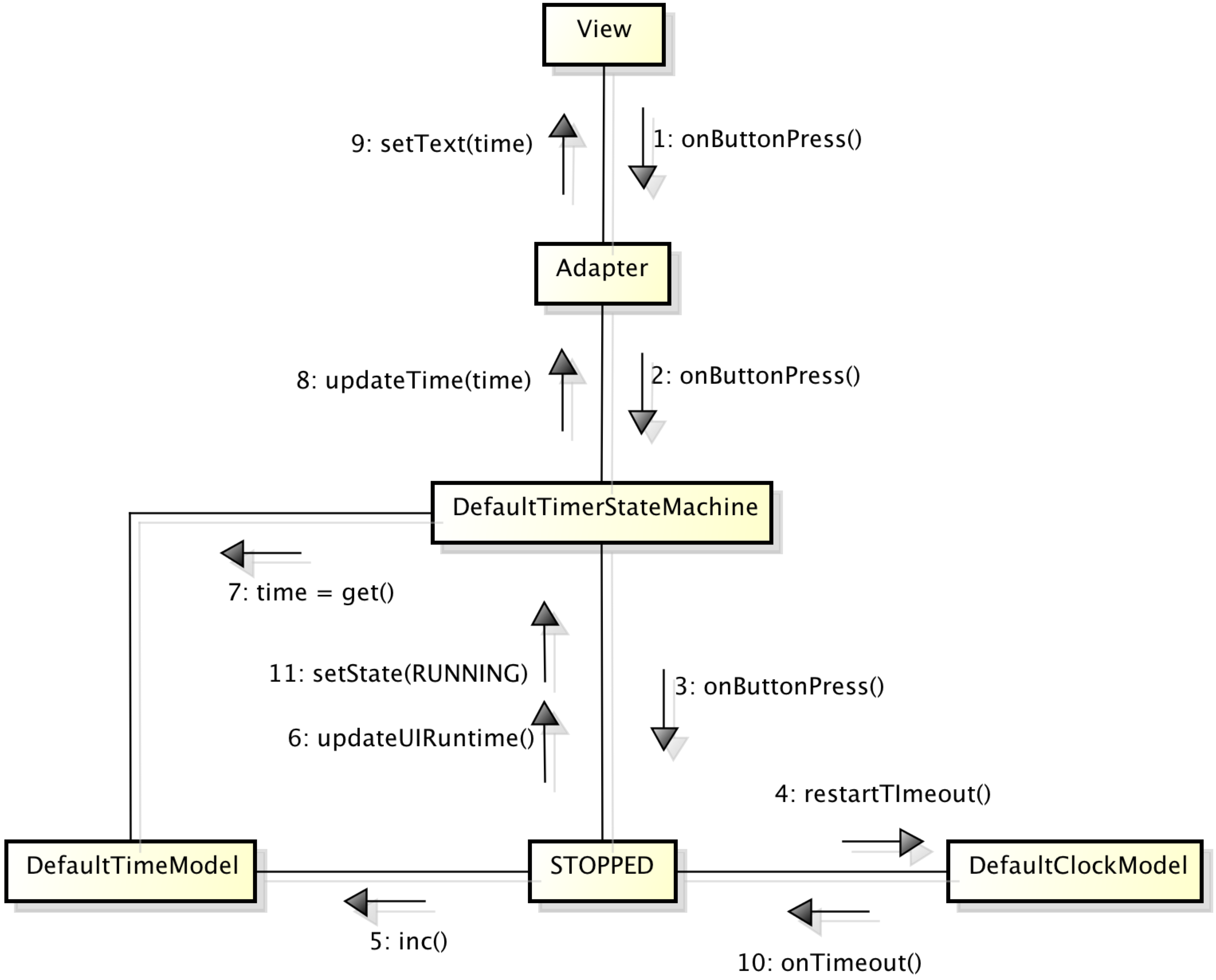}
\end{center}
\caption{Countdown timer: user input scenario (button press)}
\label{fig:CountdownTimerMVACollab1}
\end{figure}

The user input scenario in figure~\ref{fig:CountdownTimerMVACollab1}
illustrates the system's end-to-end response to a button press. The
internal timeout gets set in response to a button press. When the
timeout event actually occurs, corresponding to an invocation of the
\verb+onTimeout+ method, the system responds by transitioning to the running
state.

\begin{figure}
\begin{center}
\includegraphics[height=0.4\textheight]{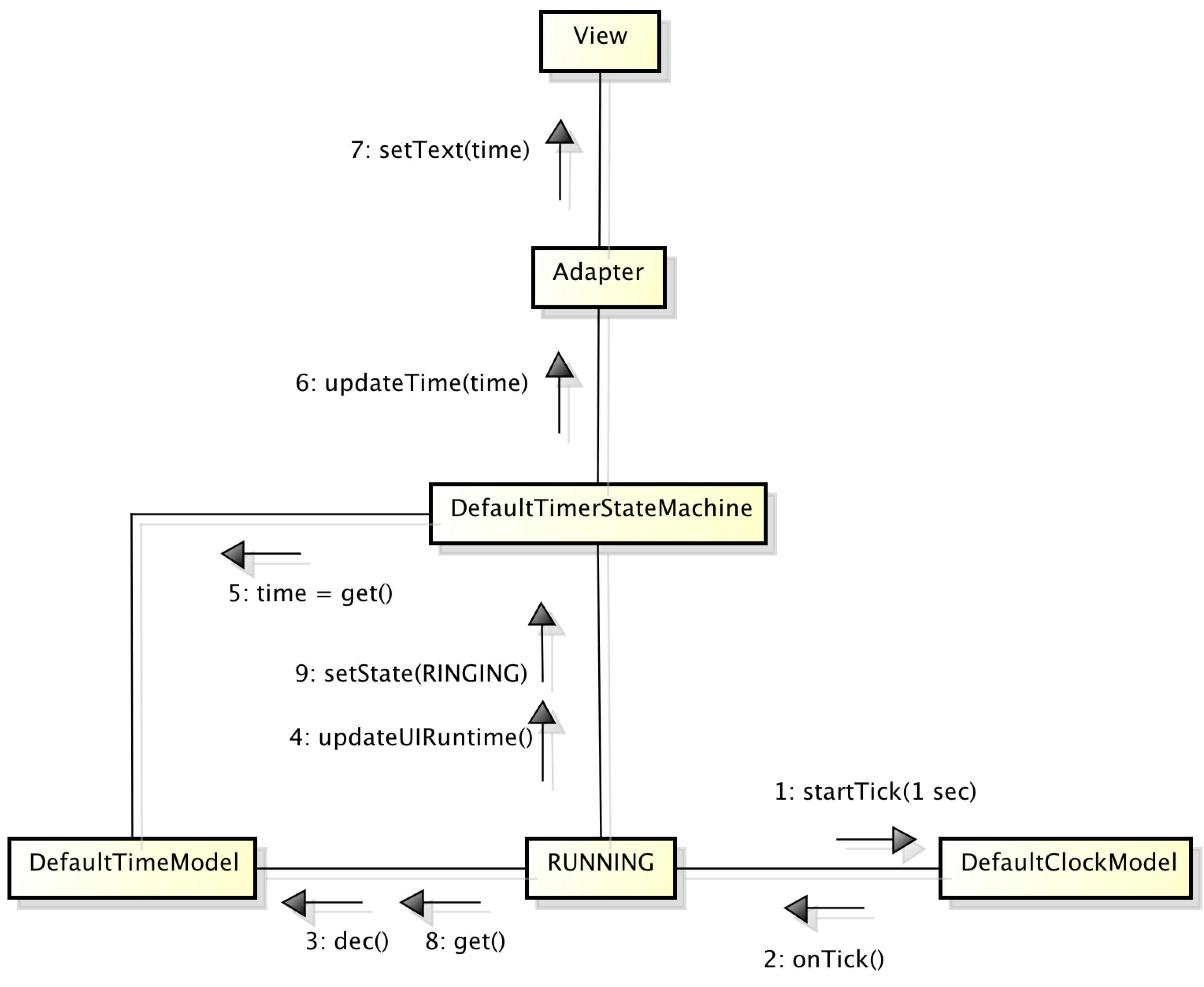}
\end{center}
\caption{Countdown timer: autonomous scenario (timeout)}
\label{fig:CountdownTimerMVACollab2}
\end{figure}

By contrast, the autonomous scenario in
figure~\ref{fig:CountdownTimerMVACollab2} shows the system's
end-to-end response to a recurring internal clock tick, corresponding
to an invocation of the \verb+onTick+ method. When the remaining time reaches
zero, the system responds by transitioning to the alarm-ringing state.

\subsection{Testing GUI applications with complex behavior and structure}

As we develop more complex applications, we increasingly benefit from
thorough automated testing. In particular, there are different
structural levels of testing: component-level unit testing,
integration testing, and system testing. Testing is particularly
important in the presence of concurrency, where timing and
nondeterminism are of concern.

In addition, as our application grows in complexity, so does our test
code, so it makes sense to use good software engineering practice in
the development of our test code. Accordingly, software design
patterns for test code have emerged, such as the 
Testclass Superclass pattern~\cite{XUnitPatterns} 
we use in section~\ref{sec:Events}.

\subsubsection{Unit-testing passive model components}

The time model is a simple passive component, so we can test it very
similarly as the bounded counter model in section~\ref{sec:Events}.

\subsubsection{Unit-testing components with autonomous behavior}

Testing components with autonomous behavior is more challenging
because we have to attach some kind of probe to observe the behavior
while taking into account the presence of additional threads.

Let's try this on our clock model. The following test verifies that a
stopped clock does not emit any tick events. 
\begin{lstlisting}
@Test
public void testStopped() throws InterruptedException { 
  final AtomicInteger i = new AtomicInteger(0); 
  model.setClockListener(new ClockListener() {
    @Override public void onTick() { i.incrementAndGet(); }
    @Override public void onTimeout() { }
  });
  Thread.sleep(5500);
  assertEquals(0, i.get());
}
\end{lstlisting}

And this one verifies that a running clock emits roughly one tick event per second.
\begin{lstlisting}
@Test
public void testRunning() throws InterruptedException { 
  final AtomicInteger i = new AtomicInteger(0); 
  model.setClockListener(new ClockListener() {
    @Override public void onTick() { i.incrementAndGet(); }
    @Override public void onTimeout() { } 
  });
  model.startTick(1 /* second */);
  Thread.sleep(5500);
  model.stopTick();
  assertEquals(5, i.get());
}
\end{lstlisting}

Because the clock model has its own timer thread, separate from the
main thread executing the tests, we need to use a thread-safe
\verb+AtomicInteger+ to keep track of the number of clock ticks across the
two threads.

\subsubsection{Unit-testing components with autonomous behavior and
  complex dependencies}

Some model components have complex dependencies that pose additional
difficulties with respect to testing. Our timer's state machine model,
e.g., expects implementations of the interfaces \verb+TimeModel+, \verb+ClockModel+,
and \verb+TimerUIUpdateListener+ to be present. We can achieve this by
manually implementing a so-called \emph{mock object}\footnote{
There are also various mocking frameworks, such as Mockito and
JMockit, which can automatically generate mock objects that represent
component dependencies from interfaces and provide APIs or
domain-specific languages for specifying test expectations.}
that unifies these three
dependencies of the timer state machine model, corresponding to the
three interfaces this mock object implements.
\begin{lstlisting}
class UnifiedMockDependency implements TimeModel, ClockModel, TimerUIUpdateListener {

  private int timeValue = -1, stateId = -1;
  private int runningTime = -1, idleTime = -1;
  private boolean started = false, ringing = false;

  public int     getTime()   { return timeValue; }
  public int     getState()  { return stateId; }
  public boolean isStarted() { return started; }
  public boolean isRinging() { return ringing; }

  @Override public void updateTime(final int tv) { this.timeValue = tv; }
  @Override public void updateState(final int stateId) { this.stateId = stateId; }
  @Override public void ringAlarm(final boolean b) { ringing = b; }

  @Override public void setClockListener(final ClockListener listener) {
    throw new UnsupportedOperationException();
  }
  @Override public void startTick(final int period) { started = true; }
  @Override public void stopTick() { started = false; }
  @Override public void restartTimeout(final int period) { }

  @Override public void reset() { runningTime = 0; }
  @Override public void inc()   { if (runningTime != 99) { runningTime++; } }
  @Override public void dec()   { if (runningTime != 0) { runningTime--; } }
  @Override public int  get()   { return runningTime; }
}
\end{lstlisting}

The instance variables and corresponding getter methods enable us to test
whether the SUT produced the expected state changes in the mock
object. The three remaining blocks of methods correspond to the three
implemented interfaces, respectively.

Now we can write tests to verify actual scenarios. In the following
scenario, we start with time 0, press the button once, expect time 1,
press the button 198 times (the max time is 99), expect time 99,
produce a timeout event, check if running, wait 50 seconds, expect
time 49 (99-50), wait 49 seconds, expect time 0, check if ringing,
wait 3 more seconds (just in case), check if still ringing, press the
button to stop the ringing, and make sure the ringing has stopped and
we are in the stopped state.

\begin{lstlisting}
@Test
public void testScenarioRun2() {
  assertEquals(R.string.STOPPED, dependency.getState());
  model.onButtonPress();
  assertTimeEquals(1);
  assertEquals(R.string.STOPPED, dependency.getState());
  onButtonRepeat(MAX_TIME * 2);
  assertTimeEquals(MAX_TIME);
  model.onTimeout();
  assertEquals(R.string.RUNNING, dependency.getState());
  onTickRepeat(50);
  assertTimeEquals(MAX_TIME - 50);
  onTickRepeat(49);
  assertTimeEquals(0);
  assertEquals(R.string.RINGING, dependency.getState());
  assertTrue(dependency.isRinging());
  onTickRepeat(3);
  assertEquals(R.string.RINGING, dependency.getState());
  assertTrue(dependency.isRinging());
  model.onButtonPress();
  assertFalse(dependency.isRinging());
  assertEquals(R.string.STOPPED, dependency.getState());
}
\end{lstlisting} 

Note that this happens in \emph{``fake time''} (fast-forward mode) because we can
make the rate of the clock ticks as fast as the state machine can keep
up.

\subsubsection{Programmatic system testing of the app}

The following is a system test of the application with all of its real
component present. It verifies the following scenario in \emph{real time}:
time is 0, press button five times, expect time 5, wait 3 seconds,
expect time 5, wait 3 more seconds, expect time 2, press \emph{stopTick}
button to reset time, expect time 0. This test also includes the
effect of all state transitions as assertions.
\begin{lstlisting}
@Test
public void testScenarioRun2() throws Throwable {
  getActivity().runOnUiThread(new Runnable() { @Override public void run() {
    assertEquals(STOPPED, getStateValue());
    assertEquals(0, getDisplayedValue());
    for (int i = 0; i < 5; i++) {
      assertTrue(getButton().performClick());
    }
  }});
  runUiThreadTasks();
  getActivity().runOnUiThread(new Runnable() { @Override public void run() {
    assertEquals(5, getDisplayedValue());
  }});
  Thread.sleep(3200); // <-- do not run this in the UI thread!
  runUiThreadTasks();
  getActivity().runOnUiThread(new Runnable() { @Override public void run() {
    assertEquals(RUNNING, getStateValue());
    assertEquals(5, getDisplayedValue());
  }});
  Thread.sleep(3200);
  runUiThreadTasks();
  getActivity().runOnUiThread(new Runnable() { @Override public void run() {
    assertEquals(RUNNING, getStateValue());
    assertEquals(2, getDisplayedValue());
    assertTrue(getButton().performClick());
  }});
  runUiThreadTasks();
  getActivity().runOnUiThread(new Runnable() { @Override public void run() {
    assertEquals(STOPPED, getStateValue());
  }});
}
\end{lstlisting}
As in section~\ref{sec:Events}, we can run this test as an in-container
instrumentation test or out-of-container using a simulated environment
such as Robolectric.

During testing, our use of threading should mirror that of the SUT:
The button press events we simulate using the \verb+performClick+ method have
to run on the UI thread of the simulated environment. While the UI
thread handles these events, we use \verb+Thread.sleep+ in the main thread of
the test runner to wait in pseudo-real-time, much like the user would
wait and watch the screen update.

Robolectric queues tasks scheduled on the UI thread until it is told
to perform these. Therefore, we must invoke the \verb+runUiThreadTasks+
method \emph{before} attempting our assertions on the UI components.

\section{Keeping the User Interface Responsive with Asynchronous Activities}

In this section, we explore the issues that arise when we use a GUI to
control long-running, processor-bound activities. In particular, we'll
want to make sure the GUI stays responsive even in such scenarios and
the activity supports progress reporting and cancelation. Our running
example will be a simple app for checking whether a number is
prime. The complete code for this example is available
online~\cite{CS313PrimeChecker}.

\subsection{The functional requirements for the prime checker app}

\begin{figure}
\begin{center}
\includegraphics[height=0.4\textheight]{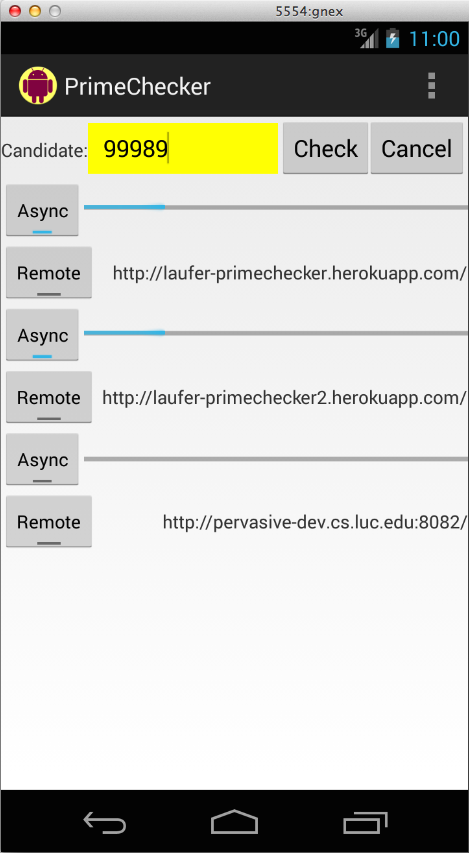}
\end{center}
\caption{Screenshot of an Android app for checking prime numbers}
\label{fig:PrimecheckerAndroid}
\end{figure}

The functional requirements for this app are as follows: 
\begin{itemize}
\item The app allows us to enter a number in a text field. 
\item When we press the \emph{check} button, the app checks whether the number we entered is prime. 
\item If we press the \emph{cancel} button, any ongoing check(s) are discontinued.
\end{itemize}
Figure~\ref{fig:PrimecheckerAndroid} shows a possible UI for this app.

To check whether a number is prime, we can use this iterative brute-force algorithm.
\begin{lstlisting}
protected boolean isPrime(final long i) {
  if (i < 2) return false;
  final long half = i / 2;
  for (long k = 2; k <= half; k += 1) {
    if (isCancelled() || i % k == 0) return false;
    publishProgress((int) (k * 100 / half));
  }
  return true;
}
\end{lstlisting}
For now, let's ignore the \verb+isCancelled+ and \verb+updateProgress+ methods and
agree to discuss their significance later in this section.

While this is not an efficient prime checker implementation, this app
will allow us to explore and discuss different ways to run one or more
such checks. In particular, the fact that the algorithm is heavily
processor-bound makes it an effective running example for discussing
whether to move such activities to the background (or remote servers).


\subsection{The problem with foreground tasks}

As a first attempt, we now can run the \verb+isPrime+ method from within our
event listener in the current thread of execution (the main GUI
thread). 
\begin{lstlisting}
  final PrimeCheckerTask t = new PrimeCheckerTask(progressBars[0], input);
  localTasks.add(t);
  t.onPreExecute();
  final boolean result = t.isPrime(number);
  t.onPostExecute(result);
  localTasks.clear();
\end{lstlisting}
The methods \verb+onPreExecute+ and \verb+onPostExecute+ are for resetting the user
interface and displaying the result.

As shown in~\ref{table:responseTimes} below, response times (in
seconds) are negligible for very small numbers but increase roughly
linearly. ``$\ll 1$'' means no noticeable delay, and ``$*$'' means that
the test was not canceled before it completed.

The actual execution targets for the app or \verb+isPrime+ implementation are 

\begin{itemize}
\item Samsung Galaxy Nexus I9250 phone (2012 model): dual-core
  1.2 GHz Cortex-A9 ARM processor with 1GB of RAM (using one core)
\item Genymotion x86 Android emulator with 1GB of RAM and one
  processor running on a MacBook Air
\item MacBook Air (mid-2013) with 1.7 GHz Intel Core i7 and 8GB of RAM
\item Heroku free plan with one web dyno with 512MB of RAM
\end{itemize}

\begin{table}
\[
\begin{array}{rrrrr}
\text{\emph{execution target}} & \text{phone} & \text{emulator} & \text{computer} & \text{web service} \\
\text{\emph{prime}} \\
1013 & \ll 1 & \ll 1 & \ll 1 & \ll 1 \\
10007 & 1 & \ll 1 & \ll 1 & \ll 1 \\
100003 & 3 & 1 & \ll 1 & \ll 1 \\
1000003 & 27 & 6 & \ll 1 &  1 \\
10000169 & * & 60 & 2 & 2 \\
100000007 & * & * & 8 & 8
\end{array}
\]	
\caption{Response times for checking different prime numbers on representative execution targets}
\label{table:responseTimes}
\end{table}

For larger numbers, the user interface on the device freezes
noticeably while the prime number check is going on, so it does not
respond to pressing the cancel button. There is no progress reporting
either: The progress bar jumps from zero to 100 when the check
finishes. In the UX (user experience) world, any freezing for more
than a fraction of a second is considered unacceptable, especially
without progress reporting.

\subsection{Reenter the single-threaded user interface model}

The behavior we are observing is a consequence of the single-threaded
execution model underlying Android and similar GUI frameworks. As
discussed above, in this design, all UI events, including user inputs
such as button presses and mouse moves, outputs such as changes to
text fields, progress bar updates, and other component repaints, and
internal timers, are processed sequentially by a single thread, known
in Android as the \emph{main thread} (or UI thread). We will say \emph{UI thread}
for greater clarity.

To process an event completely, the UI thread needs to dispatch the
event to any event listener(s) attached to the event source
component. Accordingly, single-threaded UI designs typically come with
two rules:

\begin{enumerate}
\item To ensure responsiveness, code running on the UI thread must
  never block.  
\item To ensure thread-safety, only code running on the
  UI thread is allowed to access the UI components.
\end{enumerate}

In interactive applications, running for a long time is almost as bad
as blocking indefinitely on, say, user input. To understand exactly
what is happening, let's focus on the point that events are processed
sequentially in our scenario of entering a number and attempting to
cancel the ongoing check.
\begin{itemize}
\item The user enters the number to be checked.
\item The user presses the check button.
\item To process this event, the UI thread runs the attached listener,
  which checks whether the number is prime.
\item While the UI thread running the listener, all other incoming UI
  events---pressing the cancel button, updating the progress bar,
  changing the background color of the input field, etc.---are \emph{queued}
  sequentially.
\item Once the UI thread is done running the listener, it will process
  the remaining events on the queue. At this point, the cancel button
  has no effect anymore, and we will instantly see the progress bar
  jump to 100\% and the background color of the input field change
  according to the result of the check.
\end{itemize}

So why doesn't Android simply handle incoming events concurrently,
say, each in its own thread? The main reason not to do this is that it
greatly complicates the design while at the same time sending us back
to square one in most scenarios: Because the UI components are a
shared resource, to ensure thread safety in the presence of race
conditions to access the UI, we would now have to use mutual exclusion
in every event listener that accesses a UI component. Because that is
what event listeners typically do, in practice, mutual exclusion would
amount to bringing back a sequential order. So we would have greatly
complicated the whole model without effectively increasing the extent
of concurrency in our system.

There are two main approaches to keeping the UI from freezing while a
long-running activity is going on.

\subsection{Breaking up an activity into small units of work}

The first approach is still single-threaded: We break up the
long-running activity into very small units of work to be executed
directly by the event-handling thread. When the current chunk is about
to finish, it schedules the next unit of work for execution on the
event-handling thread. Once the next unit of work runs, it first
checks whether a cancelation request has come in. If so, it simply
will not continue, otherwise it will do its work and then schedule its
successor. This approach allows other events, such as reporting
progress or pressing the cancel button, to get in between two
consecutive units of work and will keep the UI responsive as long as
each unit executes fast enough.

Now, in the same scenario as above---entering a number and attempting
to cancel the ongoing check---the behavior will be much more
responsive:

\begin{itemize}
\item The user enters the number to be checked.
\item The user presses the check button.
\item To process this event, the UI thread runs the attached listener,
  which makes a little bit of progress toward checking whether the
  number is prime and then schedules the next unit of work on the
  event queue.
\item Meanwhile, the user has pressed the cancel button, so this event
  is on the event queue \emph{before} the next unit of work toward checking
  the number.
\item Once the UI thread is done running the first (very short) unit
  of work, it will run the event listener attached to the cancel
  button, which will prevent further units of work from running.
\end{itemize}

\subsection{Asynchronous tasks to the rescue}

The second approach is typically multi-threaded: We represent the
entire activity as a separate asynchronous task. Because this is such
a common scenario, Android provides the abstract class \verb+AsyncTask+ for
this purpose.

\begin{lstlisting}
public abstract class AsyncTask<Params, Progress, Result> {
  protected void onPreExecute() { }
  protected abstract Result doInBackground(Params... params);
  protected void onProgressUpdate(Progress... values) { }
  protected void onPostExecute(Result result) { }
  protected void onCancelled(Result result) { }
  protected final void publishProgress(Progress... values) { ... }
  public final boolean isCancelled() { ... }
  public final AsyncTask<...> executeOnExecutor(Executor exec, Params... ps) { ... }
  public final boolean cancel(boolean mayInterruptIfRunning) { ... }
}
\end{lstlisting}

The three generic type parameters are \verb+Params+, the type of the
arguments of the activity; \verb+Progress+, they type of the progress values
reported while the activity runs in the background, and \verb+Result+, the
result type of the background activity. Not all three type parameters
have to be used, and we can use the type \verb+Void+ to mark a type parameter
as unused.

When an asynchronous task is executed, the task goes through the
following lifecycle:
\begin{itemize}
\item \verb+onPreExecute+ runs on the UI thread and is used to set up
  the task in a thread-safe manner.
\item \verb+doInBackground(Params...)+ is an abstract template method that we
  override to perform the desired task. Within this method, we can
  report progress using \verb+publishProgress(Progress...)+ and check for
  cancelation attempts using \verb+isCancelled()+.
\item \verb+onProgressUpdate(Progress...)+ is scheduled on the UI thread
  whenever the background task reports progress and runs whenever the
  UI thread gets to this event. Typically, we use this method to
  advance the progress bar or display progress to the user in some
  other form.
\item \verb+onPostExecute(Result)+ receives the result of the background task
  as an argument and runs on the UI thread after the background task
  finishes.
\end{itemize}

\subsection{Using AsyncTask in the prime number checker}

We set up the corresponding asynchronous task with an input of type
\verb+Long+, progress of type \verb+Integer+, and result of type \verb+Boolean+. In
addition, the task has access to the progress bar and input text field
in the Android GUI for reporting progress and results, respectively.

The centerpiece of our solution is to invoke the \verb+isPrime+ method from
the main method of the task, \verb+doInBackground+. The auxiliary methods
\verb+isCancelled+ and \verb+publishProgress+ we saw earlier in the implementation
of \verb+isPrime+ are for checking for requests to cancel the current task
and updating the progress bar, respectively. \verb+doInBackground+ and the
other lifecycle methods are implemented here:
\begin{lstlisting}
@Override protected void onPreExecute() {
  progressBar.setMax(100);
  input.setBackgroundColor(Color.YELLOW);
}

@Override protected Boolean doInBackground(final Long... params) {
  if (params.length != 1)
    throw new IllegalArgumentException("exactly one argument expected");
  return isPrime(params[0]);
}

@Override protected void onProgressUpdate(final Integer... values) {
  progressBar.setProgress(values[0]);
}

@Override protected void onPostExecute(final Boolean result) {
  input.setBackgroundColor(result ? Color.GREEN : Color.RED);
}

@Override protected void onCancelled(final Boolean result) {
  input.setBackgroundColor(Color.WHITE);
}
\end{lstlisting}

When the user presses the cancel button in the UI, any currently
running tasks are canceled using the control method \verb+cancel(boolean)+,
and subsequent invocations of \verb+isCancelled+ return \verb+false+; as a result,
the \verb+isPrime+ method returns on the next iteration.

How often to check for cancelation attempts is a matter of
experimentation: Typically, it is sufficient to check only every so
many iterations to ensure that the task can make progress on the
actual computation. Note how this design decision is closely related
to the granularity of the units of work in the single-threaded design
discussed in section~\ref{sec:SingleThreaded} above.

\subsection{Execution of asynchronous tasks in the background}

So far, we have seen how to define background tasks as subclasses of
the abstract framework class \verb+AsyncTask+. Actually executing background
tasks arises as an orthogonal concern with the following strategies to
choose from for assigning tasks to worker threads:
\begin{itemize}
\item \emph{Serial executor:} Tasks are queued and executed by a single
  background thread.
\item \emph{Thread pool executor:} Tasks are executed concurrently by a pool
  of background worker threads. The default thread pool size depend on
  the available hardware resources; a typical pool size even for a
  single-core Android device is two.
\end{itemize}

In our example, we can schedule \verb+PrimeCheckerTask+ instances on a thread pool executor:
\begin{lstlisting}
  final PrimeCheckerTask t = new PrimeCheckerTask(progressBars[i], input);
  localTasks.add(t);
  t.executeOnExecutor(AsyncTask.THREAD_POOL_EXECUTOR, number);
\end{lstlisting}

This completes the picture of moving processor-bound, potentially
long-running activities out of the UI thread but in a way that they
can still be controlled by the UI thread. 

Additional considerations apply when targeting symmetric multi-core
hardware (SMP), which is increasingly common among mobile
devices. While the application-level, coarse-grained concurrency
techniques discussed in this article still apply to multi-core
execution, SMP gives rise to more complicated low-level memory
consistency issues than those discussed above in
section~\ref{sec:ThreadSafety}. An in-depth discussion of Android app
development for SMP hardware is available here~\cite{AndroidSMP}.

\section{Summary}

In this article, we have studied various parallel and distributed
computing topics from a user-centric software development
perspective. Specifically, in the context of mobile application development,
we have studied the basic building blocks of interactive applications
in the form of events, timers, and asynchronous activities, along with
related software modeling, architecture, and design topics.

The complete source code for the examples from this article, 
along with instructions for building and running these examples,
is available from~\cite{CS313Ex}.
For further reading on designing
concurrent object-oriented software, please have a look
at~\cite{HPJPC,CPJ2E,JCIP,Magee}.

\section*{Acknowledgments}

We are grateful to our graduate students Michael Dotson and Audrey
Redovan for having contributed their countdown timer implementation,
and to our colleague Dr.\ Robert Yacobellis for providing feedback on this
article and trying these ideas in the classroom. 

We are also grateful to the anonymous CDER reviewers for their helpful
suggestions.

\bibliographystyle{alpha}
\bibliography{bib/chap02}

\end{document}